\documentclass{webofc}
\usepackage[varg]{txfonts}   
\usepackage{ulem}
\begin{document}
\title{Compact QED: the photon propagator, confinement and positivity violation for the pure gauge theory}
%
%

\author{\firstname{Orlando} \lastname{Oliveira}\inst{1}\fnsep\thanks{\email{orlando@uc.pt}} \and
        \firstname{Lee C.} \lastname{Loveridge}\inst{1,2}\fnsep\thanks{\email{loverilc@piercecollege.edu}} \and
        \firstname{Paulo J} \lastname{Silva}\inst{1}\fnsep\thanks{\email{psilva@uc.pt}}
}

\institute{CFisUC, Department of Physics, University of Coimbra, 3004-516 Coimbra, Portugal
\and
           Los Angeles Pierce College, 6201 Winnetka Ave., Woodland Hills CA 91371, USA
          }

\abstract{%
  The lattice Landau gauge photon propagator for the pure gauge theory is revisited using large lattices. For the confined case we show that it has an associated linearly growing potential, it has a mass gap, that is related to the presence of monopoles, and its spectral function violates positivity. In the deconfined phase, our simulations suggest that a free field theory is recovered in the thermodynamic limit.
  }
\maketitle
%
\section{Introduction}
\label{intro}
This contribution summarises the recent work by the authors \cite{Loveridge:2021qzs,Loveridge:2021wav,Loveridge:2022sih}
on the study of the lattice Landau gauge photon propagator using the compact formulation of QED. The interested reader aiming to
understand the details of the simulations should look into these references.

\begin{figure}[t]
\centering
\includegraphics[width=6cm,clip]{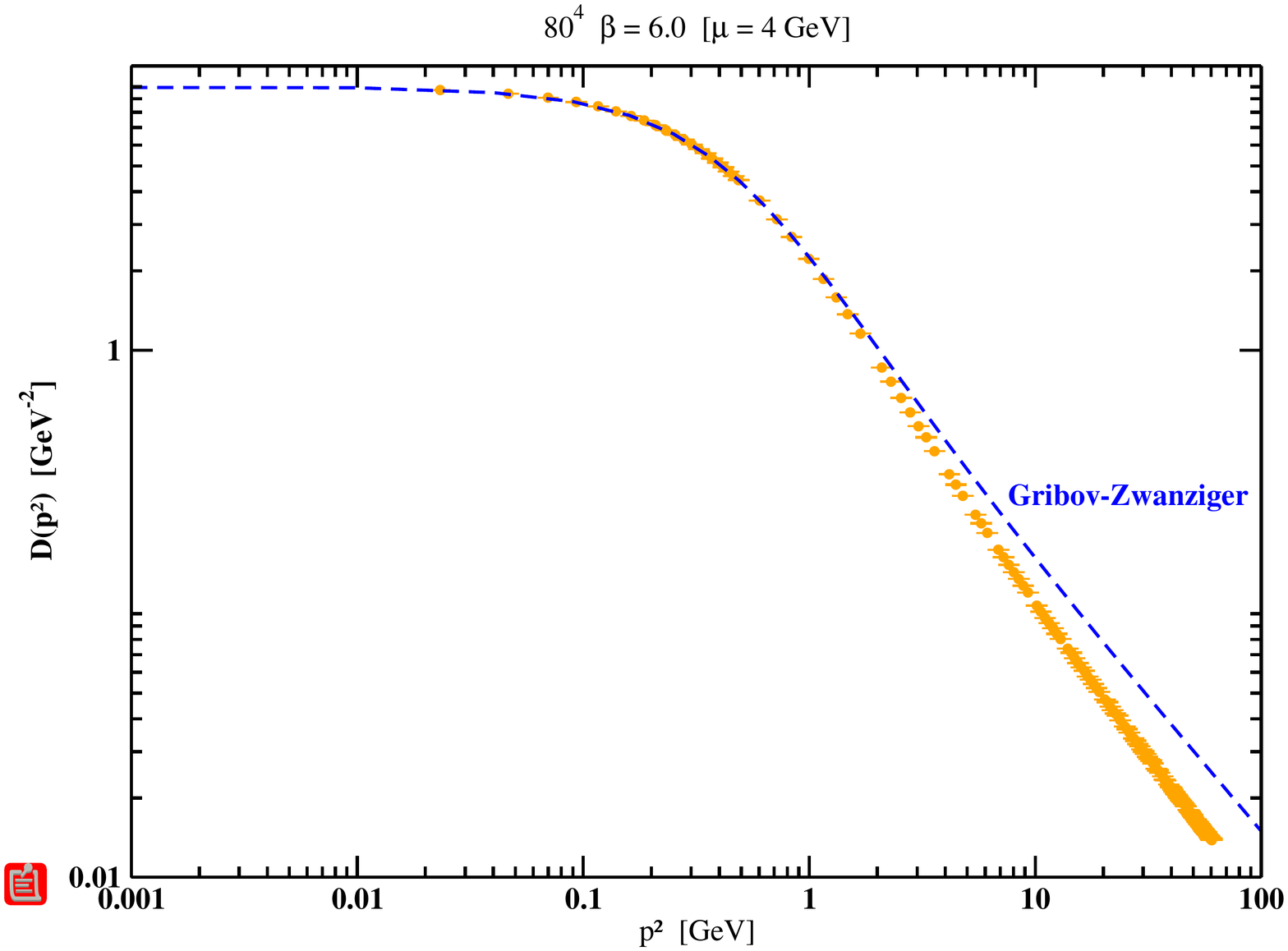} ~
\includegraphics[width=6.3cm,clip]{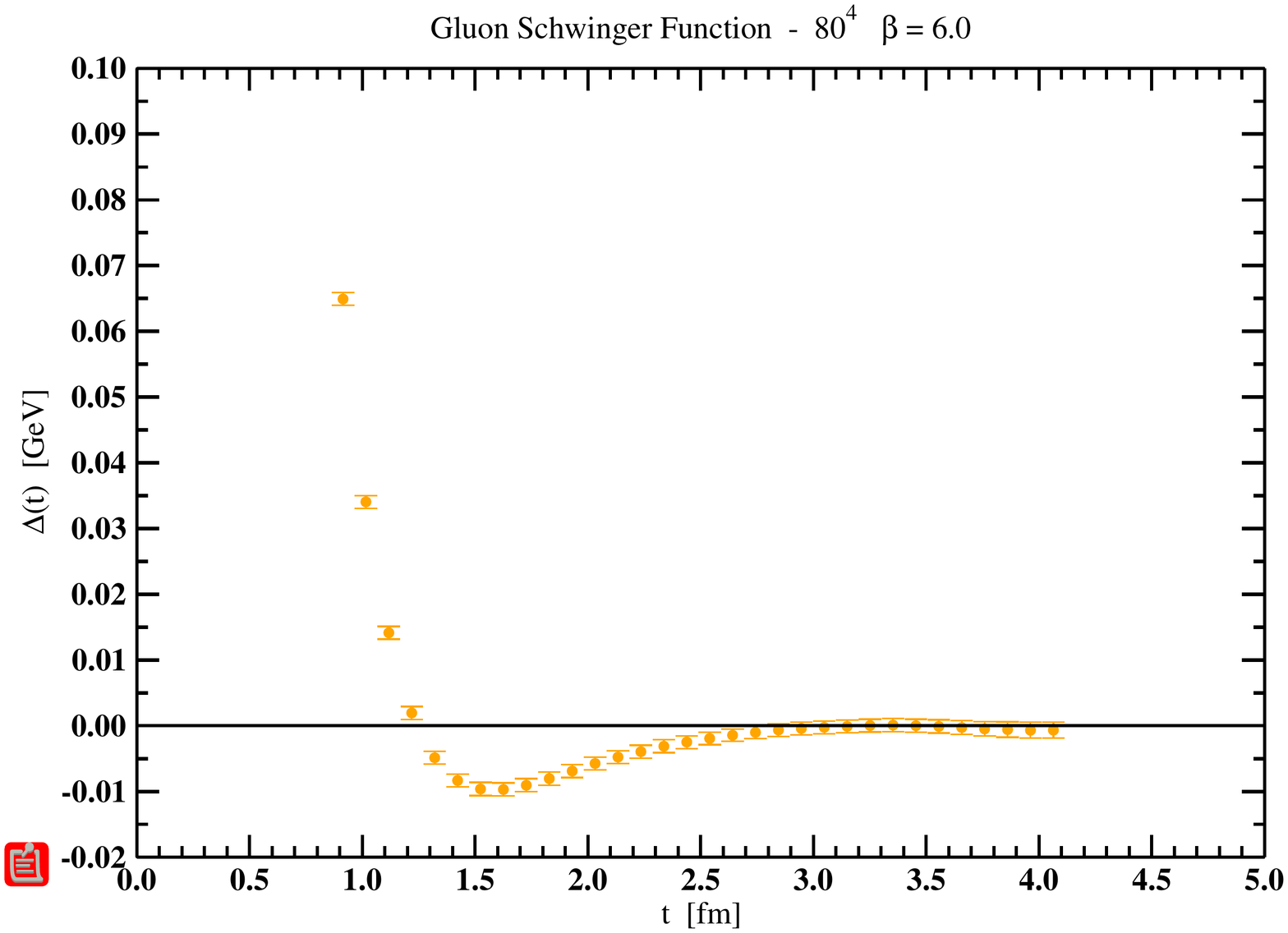}
\caption{(Left) Renormalized lattice Landau gluon propagator taken from \cite{Dudal:2018cli} renormalized in the MOM scheme at the scale $\mu = 4$ GeV and (right) the corresponding Schwinger function.}
\label{fig-gluon}       
\end{figure}

\begin{figure}[t]
\centering
\includegraphics[width=6cm,clip]{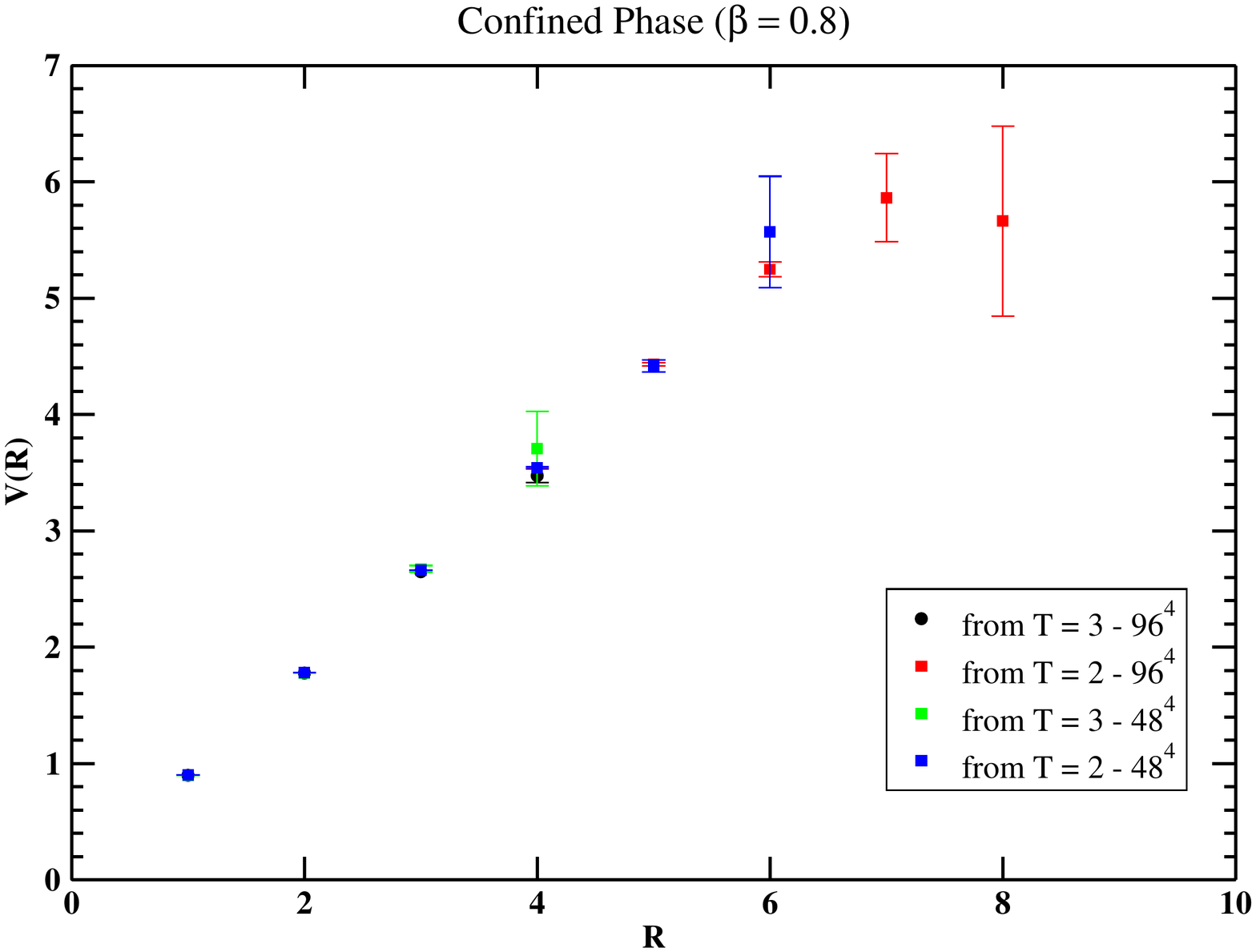} ~
\includegraphics[width=6cm,clip]{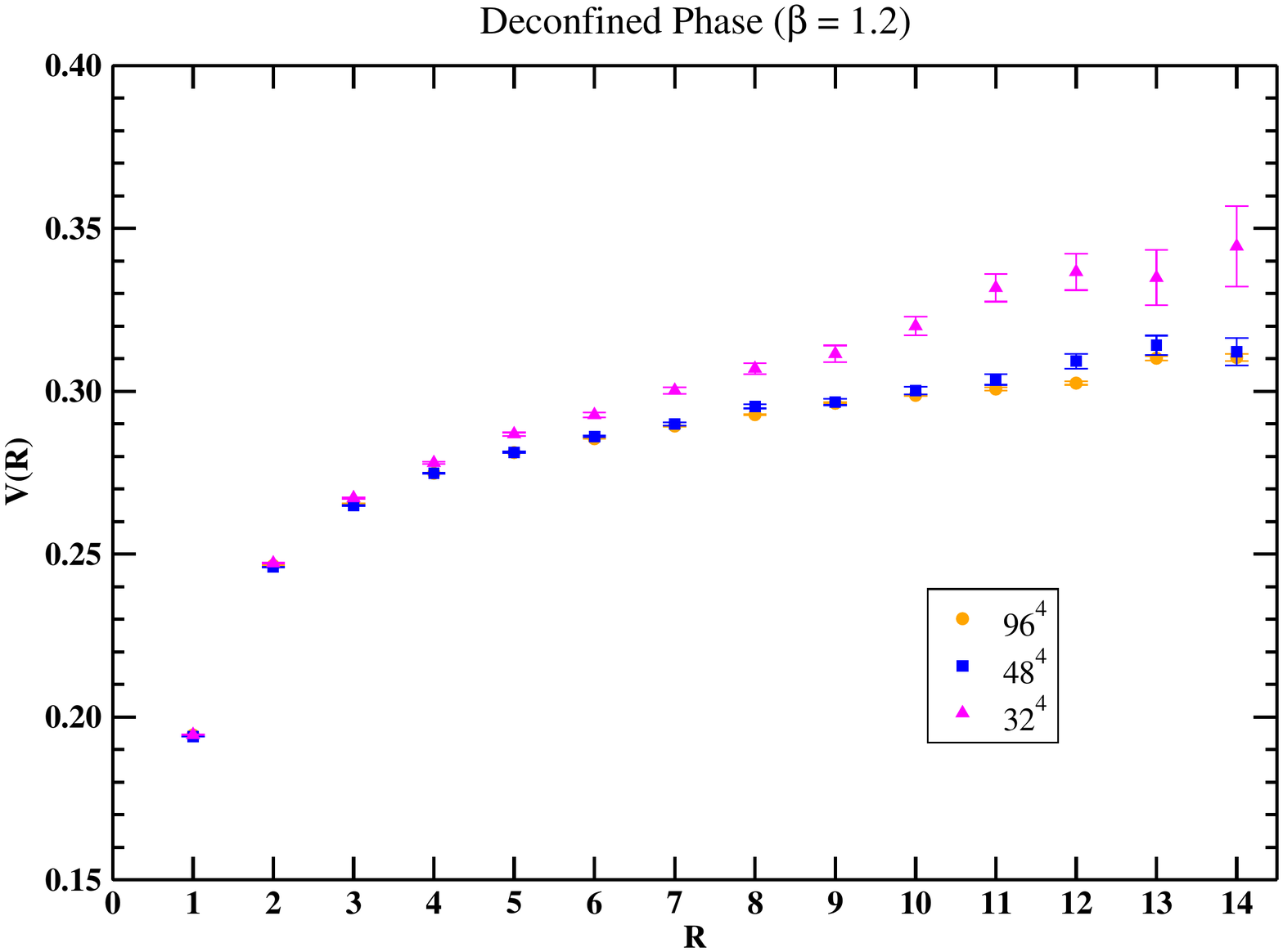}
\caption{The static potential $V(R)$ for compact QED in the confined phase (left) and deconfined phase (right). Details of the calculation can be found
in \cite{Loveridge:2021qzs}.}
\label{fig-VR}       
\end{figure}

The motivation to investigate the photon propagator comes from  QCD itself. For the pure gauge version one expects gluon confinement to take place
and that the Hilbert space of (at least) the physical states does not include single gluon states. A proof of this statement is still lacking. However,
by looking at the gluon propagator and at the corresponding Schwinger function, see Fig. \ref{fig-gluon}, tone concludes that he lattice data shows that 
the Schwinger function is not always positive definite, see \cite{Cornwall:2013zra} for a review on the positivity in QCD,  suggesting that, indeed,
gluons are confined in the sense described above. One could ask the question whether there exists another theory that can be easily simulated and that
has a phase where its quanta are confined and another phase where they are deconfined. The compact formulation of pure gauge QED on a lattice 
is such a theory. It has a single parameter $\beta = 1 /e^2$ with $e$ being the bare coupling constant. It is well known since the seventies of the last
century that for $\beta \lesssim 1$ the static potential of compact QED rises linearly with distance at large separations, see e.g. \cite{Panero:2005iu},
while for $\beta \gtrsim 1$ the static potential  becomes flat at large separations. Fig. \ref{fig-VR} taken from \cite{Loveridge:2021qzs} illustrates
the static potential for low $\beta$ and high $\beta$ values.
At least in a naive picture, this is an indication that confinement
takes place for small $\beta$ values, while the photons behave as free particles in the other phase of the theory. One should warn the reader that
the question of the continuum limit of the theory is not trivial, given that QED is not an asympotically free theory. The numerical simulations suggest
that the continuum limit is achieved by taking $\beta \rightarrow \infty$. Our results for the photon propagator suggest also, that in this limit, the theory
reproduces the usual free field theory. It is not clear what continuum theory, if any, can be associated with the low $\beta$ confining compact QED.
One can always take the point of view that, for any $\beta$, the compact formulation provides an effective theory that is defined by an (unknown)
cutoff.

Furthermore, the simulations of the compact formulation of QED allows us to access, in a relatively simple way, the topological properties of the theory.

\section{Lattice setup and expected results}
\label{sec-1}

The simulations reported here use the Wilson action
\begin{equation}
S_W = \beta \sum_x\sum_{\mu , \nu} \left( 1 - \Re \, U_{\mu\nu} (x) \right)
\end{equation}
that is given in terms of the plaquette $U_{\mu\nu}(x)$ operator built, as usual, from the link
\begin{equation}
U_\mu (x) = \exp \left\{ i \, a \, e \, A_\mu \left( x + \frac{a}{2} \hat{e}_\mu\right) \right\}
\end{equation}
where $a$ is the lattice spacing, $\hat{e}_\mu$ is the unit vector along direction $\mu$ and $A_\mu$ represents the photon field. 
From the definition of the link and of the plaquette, it follows that the phase of the plaquette $\mathcal{A}_{\mu\nu} (x)$ differs from  the sum of the
phases of the links by an integer field $m_{\mu\nu}(x)$ that represents the number of Dirac strings crossing the plaquette. Given the connection
with the Dirac strings, this integer field allows one to explore the topological properties of compact QED. 
Results on this topic can be found in \cite{Loveridge:2021qzs,Loveridge:2021wav}.

\begin{figure}[t]
\centering
\includegraphics[width=6cm,clip]{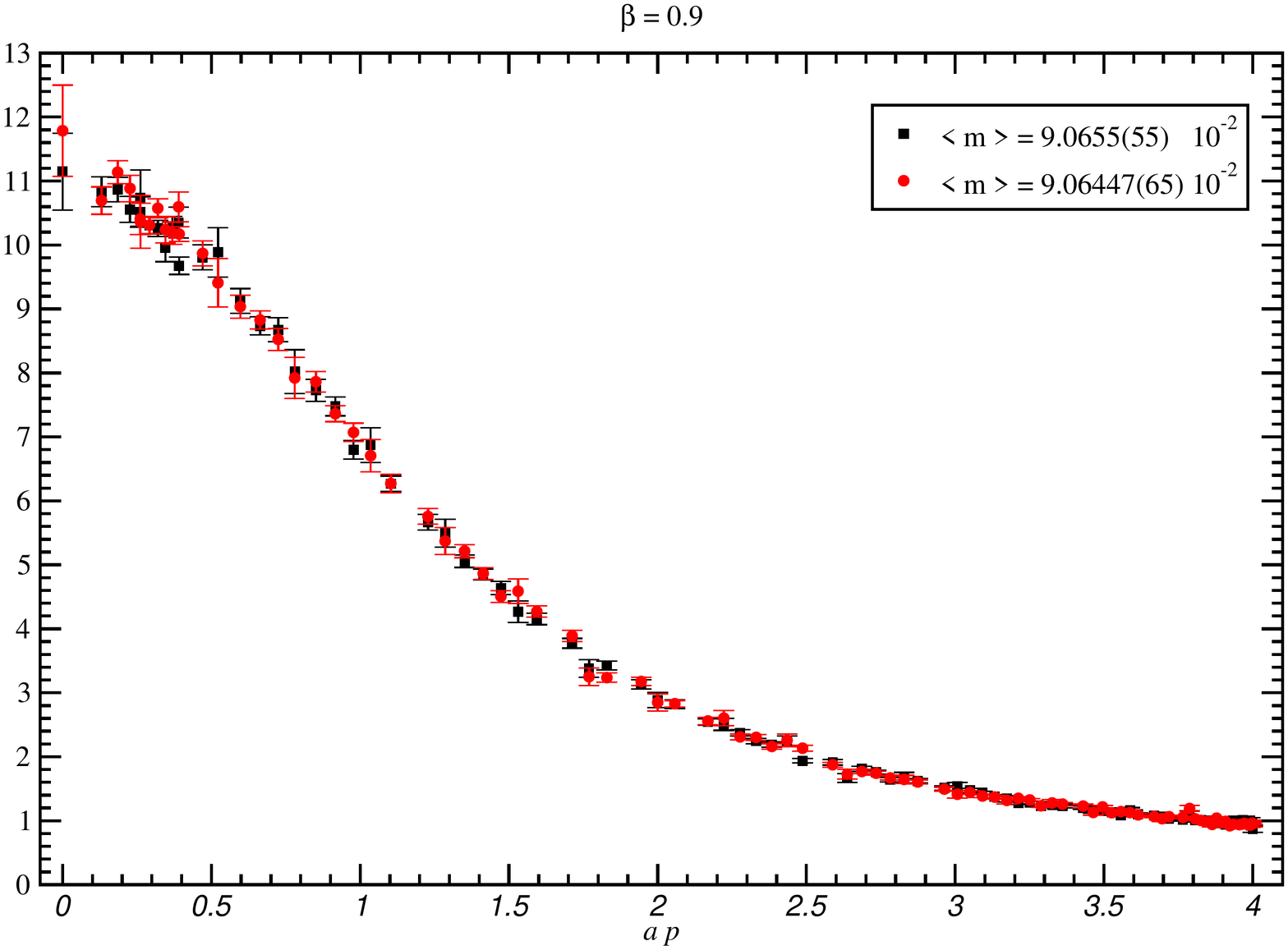} ~
\includegraphics[width=6cm,clip]{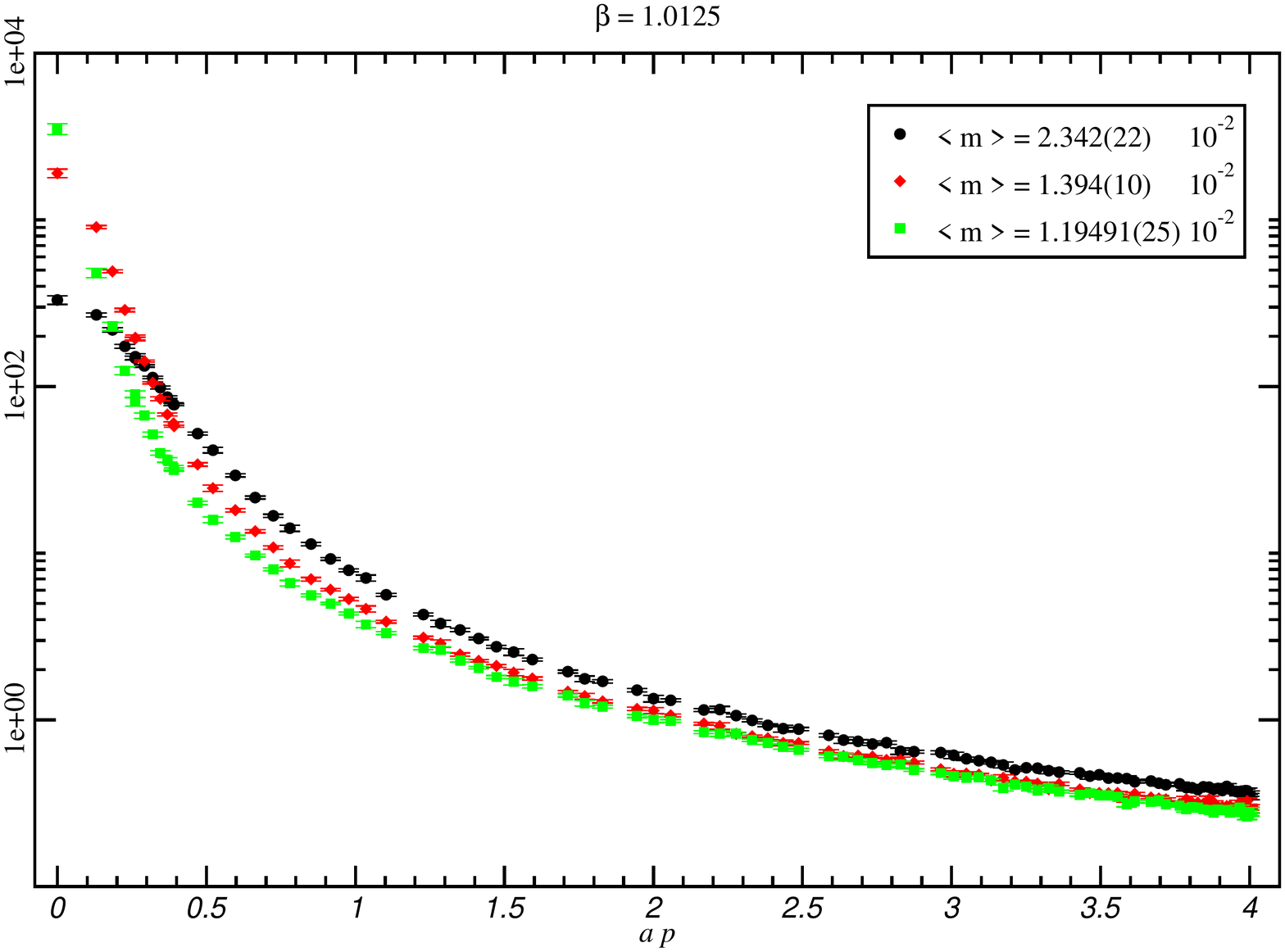} \\
\includegraphics[width=6cm,clip]{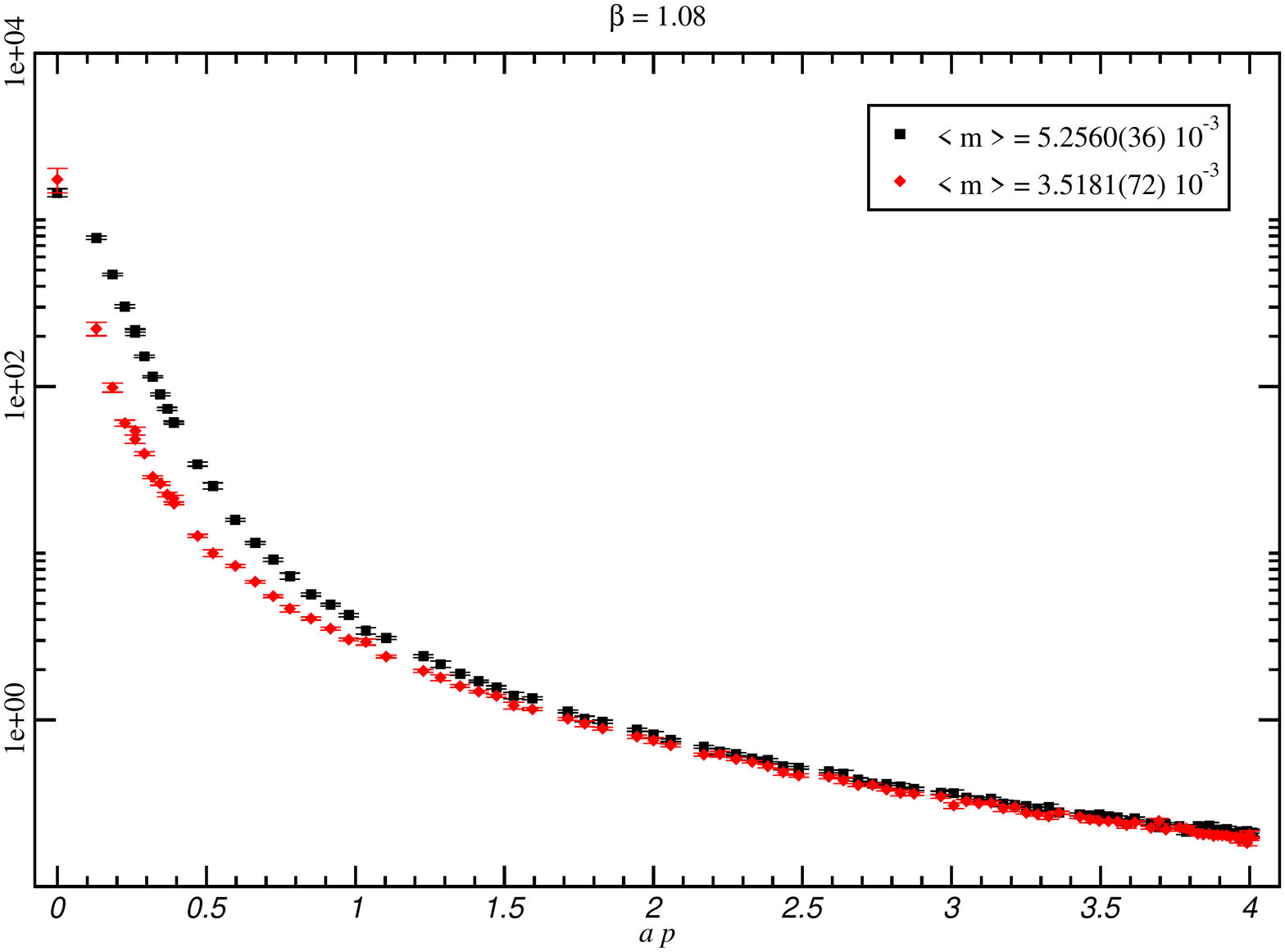} ~
\includegraphics[width=6cm,clip]{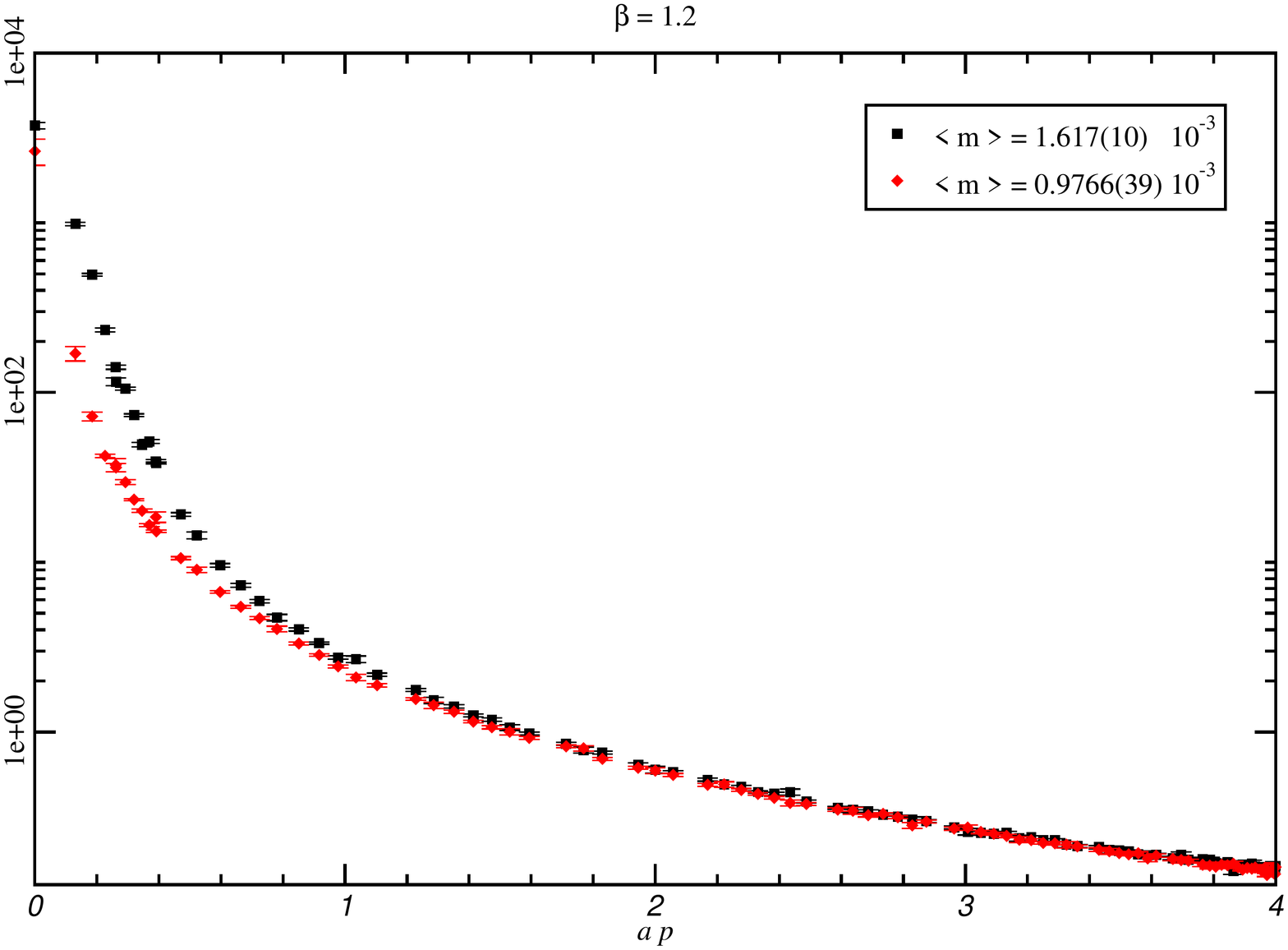} \\
\caption{The photon propagator for several $\beta$ computed from different Markov chains. $m$ stands for the average value over the lattice 
and for each configuration of $m_{\mu\nu}$. See \cite{Loveridge:2021wav} for further details.}
\label{fig-samp}       
\end{figure}

The numerical simulations for compact QED follow closely the simulations performed for non-Abelian gauge theories.
In our case, one uses the hybrid Monte Carlo (HMC) method for the sampling. It is well known that, at large $\beta$, 
the HMC method gets frozen in the topological sectors of the theory. So, to test the sampling, we have considered several Markov chains
that start from different initial configurations and compared the photon propagator that, in the Landau gauge and in momentum space, reads
\begin{equation}
    D_{\mu\nu} (p) = P^\perp_{\mu\nu} (p) ~ D(p^2) \ ,
\end{equation}
where $P^\perp_{\mu\nu}(p)$ is the transverse  momentum projection operator. We found that for the case considered and for $\beta \lesssim 1$
the propagator seems to be independent of the starting point, while for $\beta \gtrsim 1$ we observed problems in the sampling in the sense that
the propagators disagree by more than one standard deviation, see Fig. \ref{fig-samp}, and, therefore, one can expect
some quantitative dependence of the results for larger $\beta$'s.

For the definition of Landau gauge one has to take into account that given a link field, the corresponding photon field is obtained
exactly by taking the logarithm of the links, i.e. the mapping does not rely on any kind of expansion of the links as in the non-Abelian case. 
Therefore, to compute the propagator in the Landau gauge the optimizating functional that defines the gauge needs to be modified 
accordingly.

The photon propagator $D(p^2)$ should be sensitive to the phase where the simulation is performed. In the confined phase
one expects a mass gap, associated with the confinement mechanism and the non-trivial topological structure
of the theory \cite{Polyakov:1976fu}. On the other end, for large $\beta$ values one expects to recover the
perturbative propagator that, for the pure gauge theory is $D(p^2) = 1/p^2$. The simulations recover these two results, although for
the deconfined phase, due to the freezing in the sampling, the analysis has to be done with greater care.

\section{The photon propagator}
\label{sec-2}

\begin{figure}[t]
\centering
\includegraphics[width=6cm,clip]{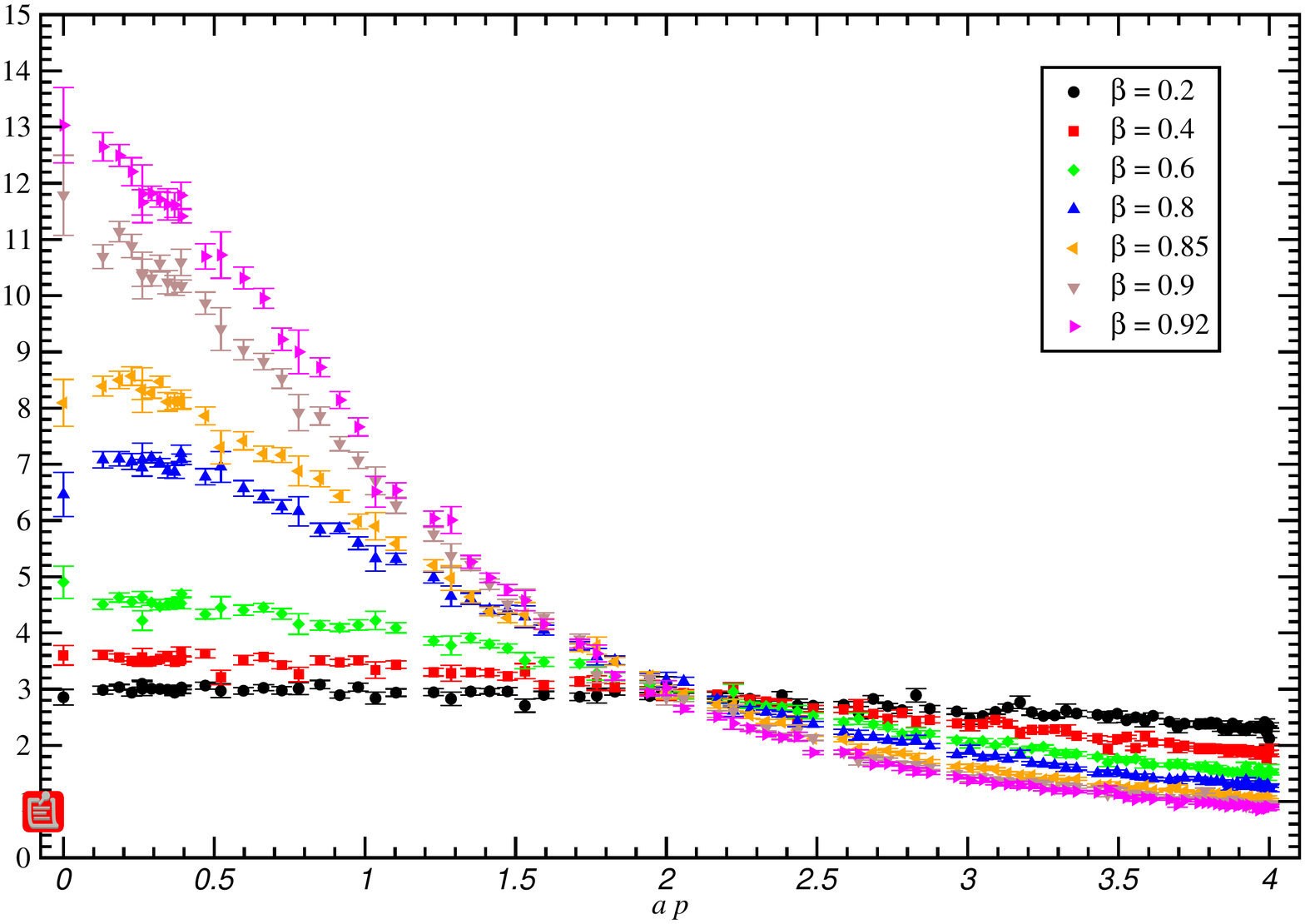} ~
\includegraphics[width=6cm,clip]{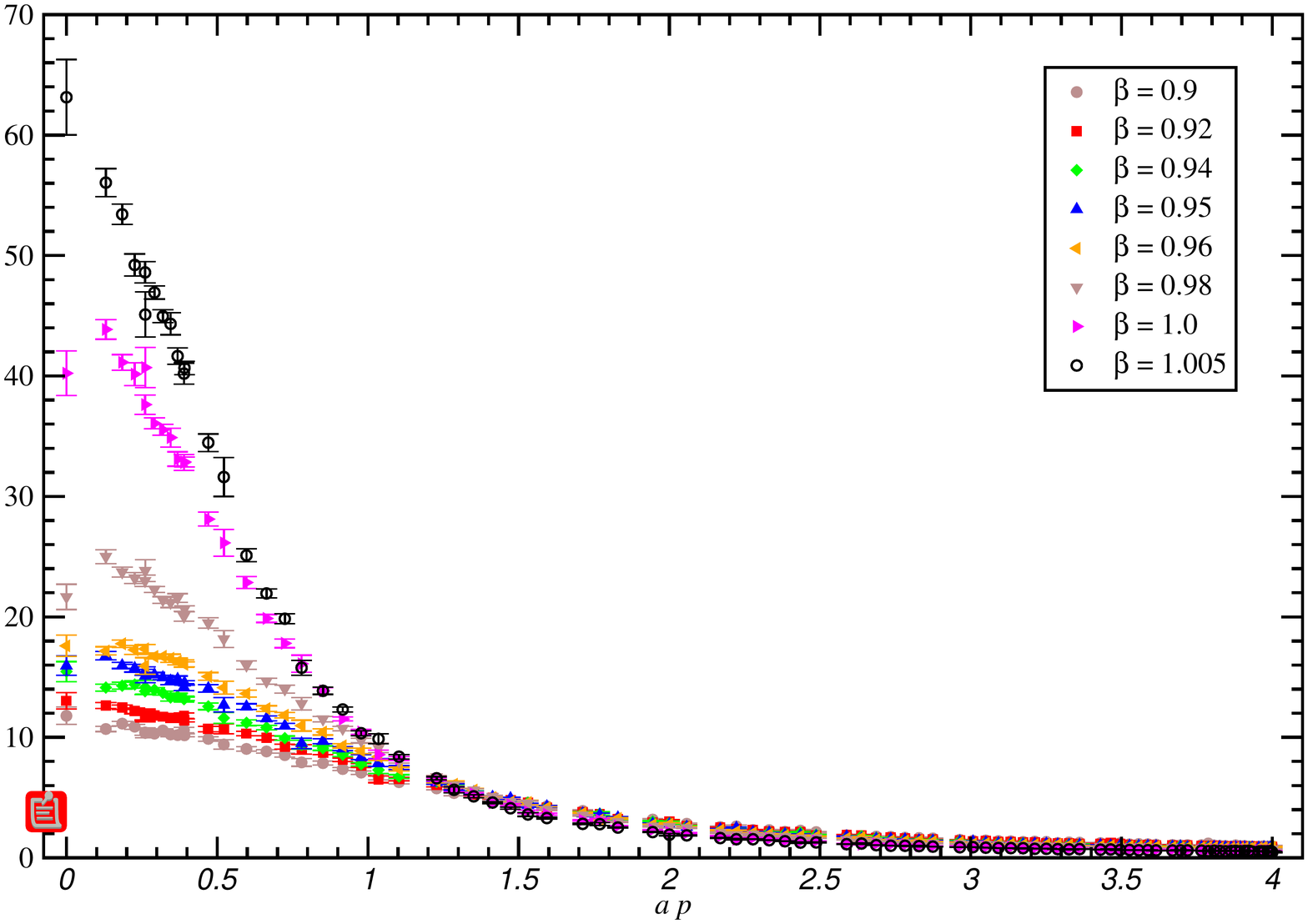} \\
\includegraphics[width=6cm,clip]{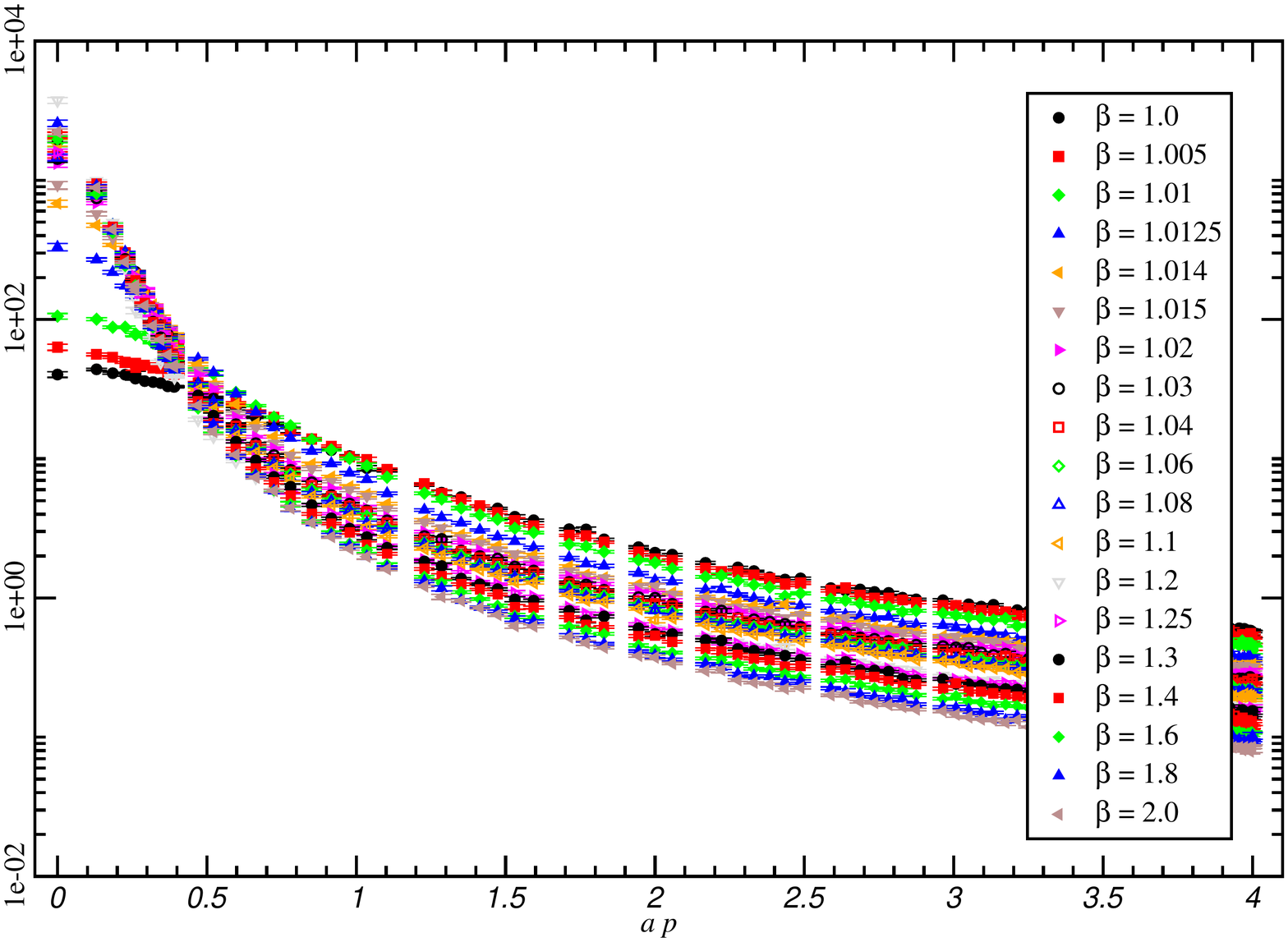} 
\caption{The bare photon propagator as a function of $\beta$.}
\label{fig-fotao}       
\end{figure}

The computation of the photon propagator follows the usual procedure. In Fig. \ref{fig-fotao} we report the bare $D(p^2)$ function
for a large set of $\beta$. Although the data is bare, it shows clearly that there is a change in the functional form of $D(p^2)$ as one crosses
$\beta \approx 1$. For compact QED there is no natural way of setting the scale to convert results into physical units. In \cite{Loveridge:2021wav}
the renormalization of the data was discussed and a procedure, that assumes a universal behaviour at high momenta, was settled. The renormalization
of the propagator data only enhances the outcome observed in the bare data.

In the confined phase where $\beta \lesssim 1$ the propagator is finite and the theory develops a mass gap. Furthermore,
by looking at the number of Dirac strings in the various configurations, one concludes that this mass gap is linked to an unequal population, away from
the zero sector, of the number of Dirac strings. Moreover, as described in our publications, for the confined phase the lattice data is compatible 
with an Yukawa type propagator $D(p^2) =  Z / (p^2 + m^2)$ allowing, from the fits, to estimate the mass gap.

On the other hand for $\beta \gtrsim 1$ the theory approaches the usual perturbative result and, in this sense, it seems to approach the continuum
limit as the lattice volume increases. It should not be forgotten that the HMC method  has ergodicity problems and there is a correlation
between the value of $m$ and the $1/p^2$ expected behaviour as can be seen in Fig. \ref{fig-samp}. The simulations show that, for the
same $\beta$, the propagators that are associated with lower values of $m$ become closer and closer to the perturbative functional form.

\section{The Schwinger function for the photon propagator}
\label{sec-3}

\begin{figure}[t]
\centering
\includegraphics[width=8cm,clip]{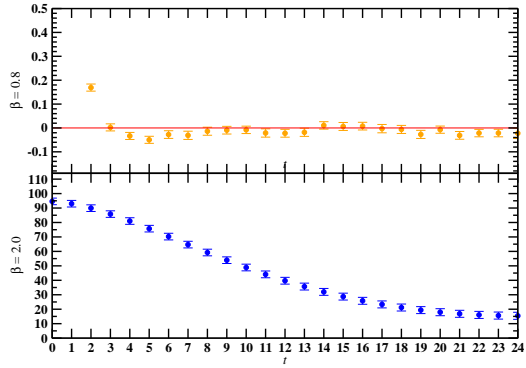}  
\caption{The photon  Schwinger function in the confined phase (upper plot) and in the deconfined phase (lower plot).}
\label{fig-Schwinger}       
\end{figure}

Assuming that the photon propagator has a spectral representation that, in Euclidean space, reads
\begin{equation}
  D(p^2) = \int^{+ \infty}_0  \, d \mu ~ \frac{ \rho ( \mu )}{p^2 + \mu} 
  \label{Eq:SpectralRepresentation}
\end{equation}
where the spectral density, for a conventional theory, is given by
\begin{equation}
  \rho ( \mu )  = \sum_n ~\delta(\mu - m^2_n) ~\left| \langle 0 | \, \mathcal{O} \, | n \rangle \right|^2 
  \label{Eq:Spectral}
\end{equation}
with the sum being over all eigenstates of the Hamiltonian and $\mathcal{O}$ stands for the operator associated with the quanta of the theory,
the photon Schwinger function is 
\begin{equation}
   C(t) = \int^{+ \infty}_0 \, \frac{dy}{2 \, \sqrt{y}} ~\rho ( y^2) ~ e^{- y \, t} 
\end{equation}
and for spectral function as in Eq. (\ref{Eq:Spectral}) $C(t)$ is always positive definite. Given that C(t) is a Laplace image of $\rho ( \mu )$, then
the measurement of $C(t)$ can test the positivity of $\rho ( \mu )$.

The photon Schwinger function in the confined and deconfined phase as seen in Fig. \ref{fig-Schwinger} shows that in the confined phase $\rho$ is not always
positive definite and, therefore, the Hilbert space of the physical states does not include a single photon state. On the other hand, for the deconfined
phase $C(t)$ is always positive definite. In this case, no statement can be made on the positivity of spectral density function.

\begin{figure}[t]
\centering
\includegraphics[width=10cm,clip]{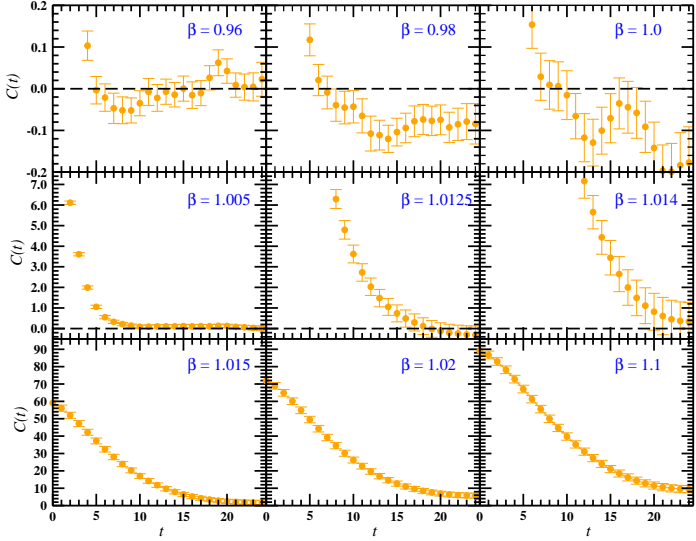}  
\caption{The photon Schwinger function knew the transition from  the confined to the deconfined phases.}
\label{fig-Schwinger2}       
\end{figure}

This result raises the question whether the non-positivity of $C(t)$ and, therefore, of $\rho$ is a characteristics of confined phase or not. Recall that for
QCD, the non-positivity of $\rho$ has been seen as an indication of the gluon confinement -- see e.g. 
\cite{Cucchieri:2004mf,Dudal:2013yva,Dudal:2019gvn} and references therein. Indeed, in perturbation theory and at one-loop the gluon
spectral density is not positive definite for all $\mu$. To help answering this question we compute $C(t)$ for compact QED
and for $\beta \simeq 1$. As seen in Fig. \ref{fig-Schwinger2}, for the confined phase $\rho$ is not positive for all $\mu$, while 
for $\beta \gtrsim 1$, it turns out that $C(t)$ does not change sign. A definitive answer on the positivity of the spectral function requires its
computation that calls for a solution of the integral equation (\ref{Eq:SpectralRepresentation}). This is an ill defined mathematical
problem that requires assuming some kind of assumption prior to access $\rho$ as e.g. in \cite{Dudal:2013yva,Dudal:2019gvn}.

\section{Summary and Conclusion}

QED is a relevant and an interesting theory that needs further studies that will certainly help in understanding confinement,
its relation to topological properties of quantum field theories and the mechanism of dynamical chiral symmetry breaking. We have
now a good understanding of the photon propagator for compact QED. From the point of view of lattice simulations it seems interesting
to perform large statistical simulations to investigate also the finite volume and finite spacing effects. Further, the observed
topological freezing in the simulations using the HMC calls for new algorithms for QED that are also of interest for
QCD simulations, see e.g. \cite{Shanahan:2018vcv,Boyda:2022nmh} and references therein.

\section*{Acknowledgements}

This work was partly supported by the FCT – Funda\c{c}\~ao para a Ci\^encia e a Tecnologia, I.P., under Projects No. UIDB/04564/2020 and UIDP/04564/2020. P. J. S. acknowledges financial support from FCT (Portugal) under Contract No. CEECIND/00488/2017. The authors acknowledge the Laboratory for Advanced Computing at the University of Coimbra (http://www.uc.pt/lca) for providing access to the HPC resource Navigator.
\bibliography{references}

\end{document}